\documentclass[doublecol]{epl2} 
\usepackage{hyperref}

\usepackage{mathrsfs}
\usepackage{xcolor}
\usepackage{amsmath}
\usepackage{amsfonts}
\usepackage{amssymb}
\usepackage{graphicx}
\usepackage{bm}
\usepackage{braket}
\usepackage{cite}
\usepackage{caption}
\usepackage{subcaption}
\usepackage[normalem]{ulem}
\title{Active particles with delayed attractions form quaking crystallites
}
\shorttitle{Active particles with delayed attractions form quaking crystallites} 
\author{Pin-Chuan Chen\inst{1} \and Klaus Kroy\inst{1} \and Frank Cichos\inst{2} \and Xiangzun Wang\inst{2} \and Viktor Holubec\inst{3}}

\shortauthor{P. C. Chen \etal}

\institute{  
\inst{1} Institute for Theoretical Physics - Universit\"at Leipzig, 04103 Leipzig, Germany
\email{chen@itp.uni-leipzig.de}
\email{klaus.kroy@uni-leipzig.de}
\\
  \inst{2} Molecular Nanophotonics Group, Peter Debye Institute for Soft Matter Physics - Universit\"at Leipzig, 04103 Leipzig, Germany
  \email{cichos@physik.uni-leipzig.de }\\ 
  \inst{3} Department of Macromolecular Physics, Faculty of Mathematics and Physics, Charles University - 18000~Prague, Czech Republic
  \email{viktor.holubec@mff.cuni.cz}
}

\abstract{Perception-reaction delays have experimentally been found  to cause a spontaneous circling of microswimmers around a targeted center. Here we investigate the many-body version of this experiment with Brownian-dynamics simulations of active particles in a plane. For short delays, the soft spherical discs form a hexagonal colloidal crystallite around a fixed target particle. Upon increasing the delay time, we observe a bifurcation to a chiral dynamical state that we can map onto that found for a single active particle. The different angular velocities at different distances from the target induce shear stresses that grow with increasing delay. As a result, tangential and, later, also radial shear bands intermittently break the rotating crystallite. Eventually, for long delays, the discs detach from the target particle to circle around it near the preferred single-particle orbit, while spinning and trembling from tidal quakes.}

\begin{document}

\maketitle

\section{Introduction}

Recent experiments with synthetic microswimmers steered toward a fixed target have revealed a spontaneous vortex formation caused by a perception-reaction delay~\cite{Wang2023}. 
The observed phenomenology can be attributed to the delay-induced aiming errors, akin to those associated with microswimmer navigation strategies employing ``vision-cone''~\cite{Bauerle2020,Loffler2021} or ``acceptance-angle''~\cite{Bregulla2014,Selmke2018} criteria. The experiment thereby established a simple paradigmatic model system for swarm forming ensembles with delayed interactions. Notably, the response of all living creatures and artificial devices to external stimuli is delayed by the time required to transfer and process information and realize the required response. 
All these systems can be classified as feedback-driven systems~\cite{Bechhoefer2005}, which are well-studied in control theory, an engineering branch of dynamical-systems theory. In physics, objects capable of active reactions to perceived stimuli, such as animals or robots, are commonly studied within the field of active matter~\cite{Ramaswamy2017}. Even though the models of active matter usually neglect perception-reaction delays, it was shown in several pioneering studies that delays can significantly impact stability, dynamical phases, and even finite-size scaling in active matter systems~\cite{sun2014time, giuggioli2015delayed,mijalkov2016engineering,leyman2018tuning,scholz2018inertial,khadka2018active,piwowarczyk2019influence,muinos2021reinforcement,holubec2021finite}. 

In this Letter, we extend the  experimental model system of Ref.~\cite{Wang2023} to system sizes that are currently inaccessible to the experimental techniques employed in~\cite{Franzl2021}. Using Brownian dynamics simulations, we find that the average angular velocity of the system still exhibits the bifurcation described in~\cite{Wang2023}, but that the many-body dynamics undergoes a surprisingly rich series of delay-induced dynamical phase transitions. For short delays, the system forms a densely packed crystallite around the target, which can be interpreted as a variant of motility-induced phase separation~\cite{Bialke2013}, with a strongly depleted gas phase. As the delay increases, the crystallite is intermittently broken up by delay-induced shear bands.  

Even for experimentally realistic noise intensities, the phenomenology observed in our simulations resembles the behavior of sheared low-temperature colloidal suspensions or athermal granular materials~\cite{Lherminier2019,Tsai2021}. An important feature of densely packed crystalline and amorphous particle assemblies is that they can only be sheared if the  packing structure is somewhat dilated to allow the particles to escape from their nearest-neighbor cages and move around each other. A typical defect structures  observed under such conditions are therefore shear bands \cite{Schall2010, Wu2009}. In the field of granular rheology, one also speaks of the dilatancy effect. It is responsible for normal stresses and the non-affine response to shear. In everyday life, you may experience it in the form of drained halos  around your feet when you step on wet sand.
In contrast to common granular and colloidal rheological setups, the shear stresses in our active-Brownian-particle ensembles are however not induced by a moving background solvent or a system boundary or immersed probe particle, but solely by the individual  particles' activity, itself. This entails some counter-intuitive consequences. Most importantly, the time delay only entails relevant navigational aiming errors if the particles are actually moving, but not if they are jammed up in a dense cluster. 
This somewhat unconventional property distinguishes our setup from the myriad of superficially related rheological problems documented in the literature. It also impedes attempts to provide a complete mechanistic interpretation of the unique succession of dynamical phases and phase transitions, described in the following.  


\section{Model} 
\begin{figure}
    \centering
    \includegraphics[width=0.9\columnwidth]{"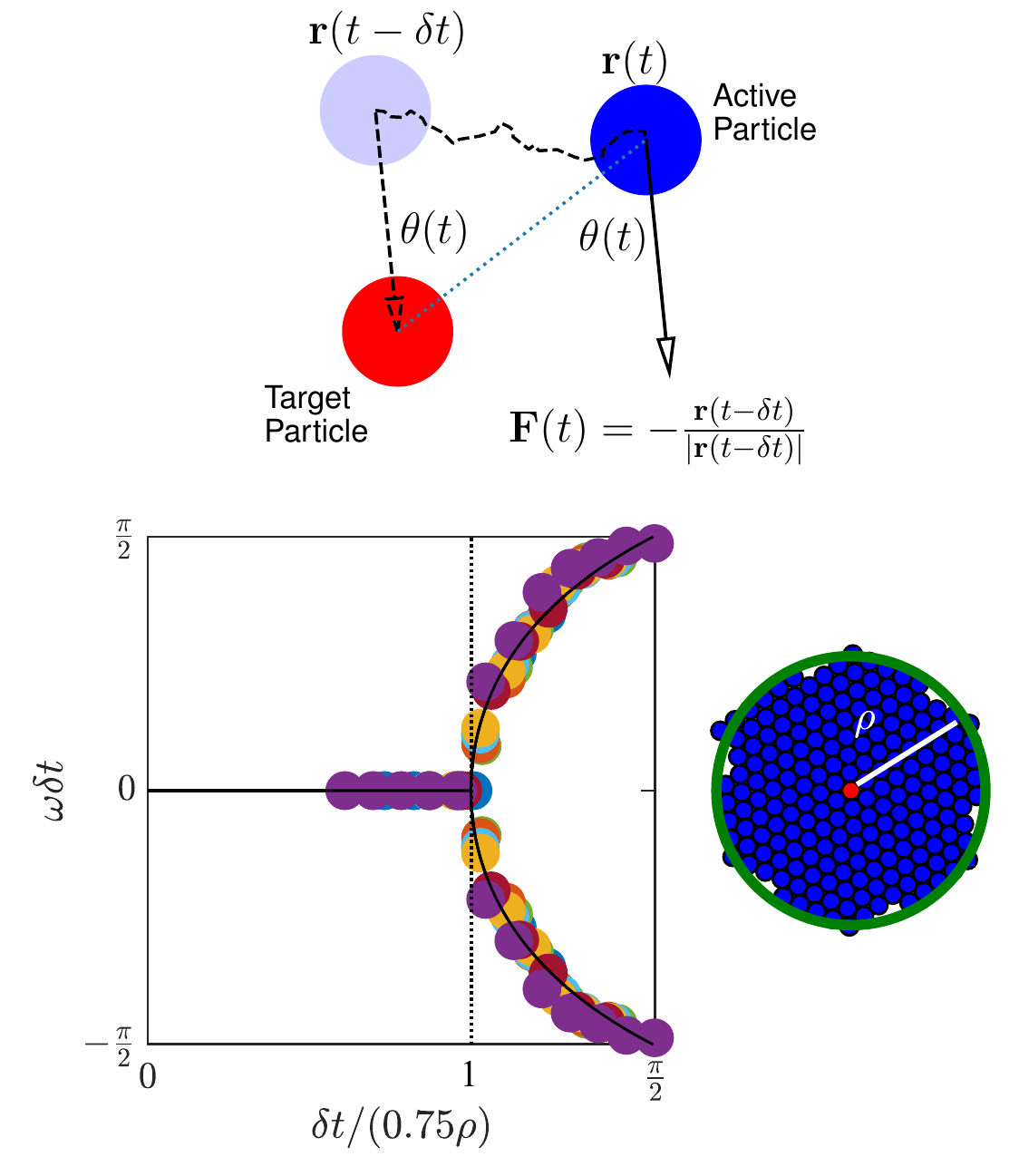"}
    \caption{Active Brownian particles (blue), swimming at constant speed, aim at a central target (red) with a perception-reaction delay $\delta t$. a) That the actual swim direction at time $t$ is determined by position $\mathbf{r}(t-\delta t)$ at the earlier time $t-\delta t$, gives rise to aiming errors and ensuing dynamical phases. b) The bifurcation diagram shows the average angular displacement $\omega \delta t$ per delay time $\delta t$. Upon increasing delay, the isotropic static phase (I) gives way to radially symmetric chiral phases (II-IV). For the yin-yang and blob phases (V and VI), $\omega \delta t \approx \pi/2$. The colors code for various particle numbers $N=30 \dots 1000$. The diagram obtained for athermal motion (diffusivity $D= 0$) remains unchanged for an experimentally realistic noise intensity ($D = 0.0136$). c) Close-packed crystallites of $N\gtrsim 30$ particles have radius $\rho = \sqrt{(N+1)/3.62}$ in units of the (soft-)particle diameter.}
    \label{fig:principle sketch}
\end{figure}

We consider a two-dimensional system of $N$ overdamped active Brownian discs, interacting via soft steric interactions. One particle is held fixed at the origin. The $N$ mobile particles aim to swim toward it with a constant speed $v_0$. As shown in Fig.~\ref{fig:principle sketch}a, they cannot react instantaneously to the detected target position, but only after a certain delay time $\delta t$.  Since the particles keep moving during this time, the resulting retarded attraction to the central target acquires important aiming errors.

We fix length and time scales by setting the particle diameter and the swim speed to unity. The dimensionless position vector $\mathbf{r}_i$ of the $i$th Brownian particle obeys the Langevin equation
\begin{equation}
        \dot{\mathbf{r}}_i(t)= \mathbf{F}_i(t)
       + k\sum_{j\neq i}\mathbf{r}_{ij}(t)\Theta\left[1-|\mathbf{r}_{ij}(t)|\right]
       +\sqrt{2D} \bm{\eta}_i(t)\,,
       \label{eq:EOM}
\end{equation}
where 
$\mathbf{F}_i(t) = -\mathbf{r}_i(t-\delta t)/|\mathbf{r}_i(t-\delta t)|$ are the intended (or nominal) velocities of the individual particles, and $t$ is the dimensionless time. The soft steric repulsion has a strength of given by the dimensionless stiffness $k$ and a range cutoff at $|\mathbf{r}_{ij}(t)|=1$, imposed by the Heaviside $\Theta$ function. The diffusivity $D$ controls the intensity of mutually independent Gaussian white noise vectors $\bm{\eta}_i$, $i=1,\dots, N$, with zero mean, $\left<\bm{\eta}_i(t)\right>=\bm{0}$, and covariance $\left<[\bm{\eta}_i(t)]_x[\bm{\eta}_j(t')]_y\right>=\delta_{ij}\delta_{xy}\delta(t-t')$. 

We studied the model for particle numbers $N =15\dots 1000$ that are neither analytically tractable nor currently realizable in experiments. The dynamical equations are solved by Brownian dynamics simulations with time step ${\rm d}t=0.001$, $k=101.4$, and $D=0.0136$. These parameters are motivated by typical experimental conditions in aqueous solvents at room temperature, if one identifies the particle diameter with  $2.19\times10^{-6}$~m and the propulsion speed with $2.16 \times 10^{-6}$~m/s~\cite{Wang2023}.
We initialized the particles randomly around the origin, let them diffuse for a time $t=\delta t$, and simulated long enough such that the system relaxed to a steady state (see the supplementary videos SM). Afterward, we continued the simulation and collected the data.
Varying $k$ and $D$ in the dynamical equations~\eqref{eq:EOM} within an experimentally reasonable range does not change the qualitative results. Hence, the relevant control parameters are the time delay $\delta t$ and particle number $N$, or the corresponding radius $\rho(N)=\sqrt{(N+1)/3.62}$ of a close packed hexagonal crystallite (see Fig.~\ref{fig:principle sketch}c). 

\section{Bifurcation}

As shown in Ref.~\cite{Wang2023}, for $N=1$, the average angular velocity $\omega$ of the single active Brownian particle around the fixed target is determined by a transcendental self-consistency equation. If the active-particle and  target diameters are set to unity, it takes the form of the ``sine map'' $\omega =  \sin (\omega \delta t)$. It exhibits a bifurcation from $\omega=0$ to $\omega \neq 0$ at $\delta t = 1$ (or, in the dimensional units of Ref.~\cite{Wang2023}, $v_0\delta t = 2a$). For $1<\delta t < \pi/2$, the single active Brownian particle ``slides'' around the target, and thus its dimensionless orbit radius is close to 1. When $\delta t > \pi/2$, swimmer and target particle lose touch and the circular orbit ``takes off''. Its radius $R = 2\delta t/\pi$ is now determined by the condition that the angular displacement of the particle per one delay time, $\omega \delta t$, is $\pi/2$. In other words, for large delay times, the particle always propels tangentially (at a right angle) to the target, corresponding to a self-selected circular orbit.  

 \begin{figure}
    \centering
\includegraphics[width=1\columnwidth]{"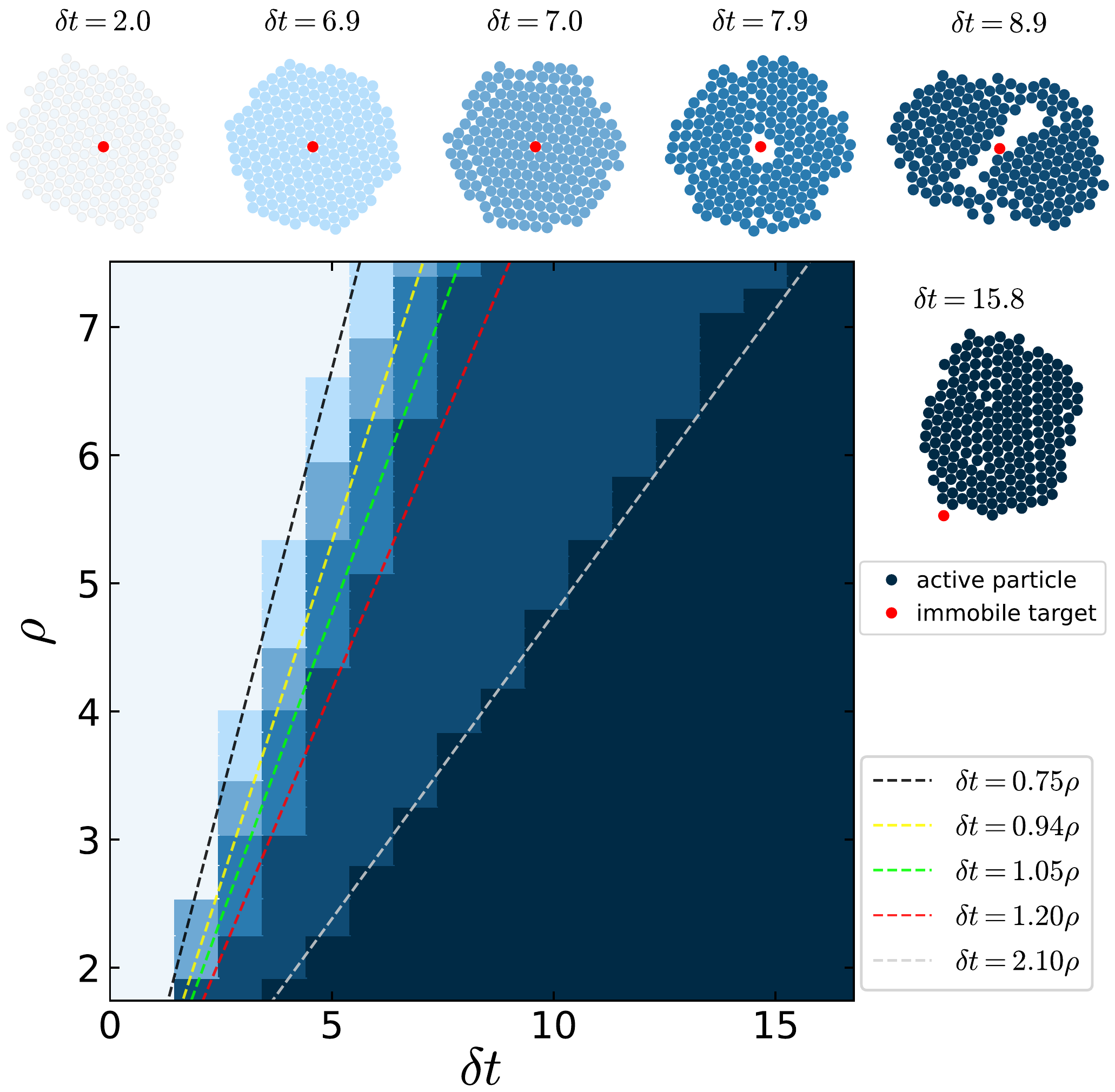"}
    \caption{
    Dynamic phase diagram. Like the preferred single-particle orbit $R$, the (binned) crystallite radius $\rho$ grows with increasing delay time $\delta t$. We distinguish phases with a (I) static, (II) spinning, and (III) quaking crystallite, and a (IV) ring, (V) yin-yang/blobs, and (VI) satellite, respectively. Notice the appearance of predominantly concentric (III), radial (IV) and criss-crossing (V-VI) shear bands that intermittently break the crystallite, giving rise to a staircase-like increase of the  shear strain $\Gamma(t)$ (3rd row of Fig.~\ref{fig:shell_analysis}), for all but the first two phases. 
    }\label{fig:phase diagram}
\end{figure}

Though not accessible experimentally, Ref~\cite{Wang2023} already demonstrated by Brownian dynamics simulations that the single-particle bifurcation diagram stays meaningful for many particles up to $N =100$. The increased particle number actually stabilizes the spontaneously chosen sense of rotation against Brownian fluctuations, rendering the transient chiral symmetry breaking quasi permanent. More importantly, the increase in particle number merely renormalizes the bifurcation diagram. As shown in Fig.~\ref{fig:phase diagram}b, the average angular particle displacement $\omega \delta t$ around the fixed target particle for $N$ ranging from 30 to 1000 indeed falls on a single master curve, if plotted against $\delta t/(0.75 \rho)$, corresponding to the renormalized sine map $0.75\rho\,\omega=\sin(\omega\delta t)$. The bifurcation curve coincides with that of a single large quasi-particle of radius $(0.75 \rho - 0.5)$, rotating around the target particle of radius 0.5. In other words, the minimum radius for active rotation (originally given by the particle diameter) equals $0.75 \rho$, in the many-body case. One can speculate that the effective radius $0.75\rho$ could coincide with the crystallite's radius of gyration $\int_0^\rho dx\, 2\pi x^2/(\pi \rho^2) \approx 0.67\rho$. This is indeed not far off,  although the data is more suggestive of a matching of the radius $R=2\delta t/\pi$ of the optimal single-particle orbit with $\rho/2$. This could suggest that spinning starts when the preferred nominal velocity components of the particles inside and outside the optimal orbit cancel out. The difficulty with such interpretation is that the actually measured nominal velocity field created by the highly frustrated active particles in the bulk of a solid crystallite is, for the relevant delays, still purely central.

As shown in Fig.~\ref{fig:phase diagram} for particle numbers $N=15 \dots 200$, when the delay time $\delta t$ is increased, the particle ensemble experiences a series of abrupt dynamical changes, thereby evolving from a static hexagonal crystallite to a continuously breaking elliptic satellite droplet, circling around the target on an orbit close to that  preferred by a single active particle. Intriguingly, the average angular velocity in all these phases obeys the effective single-particle theory well. In fact, the single-particle theory can be used as a starting point for understanding most of the features of the various dynamical phases of the many-body model.

\begin{figure*}[t!]
    \centering
    \includegraphics[width=1\textwidth]{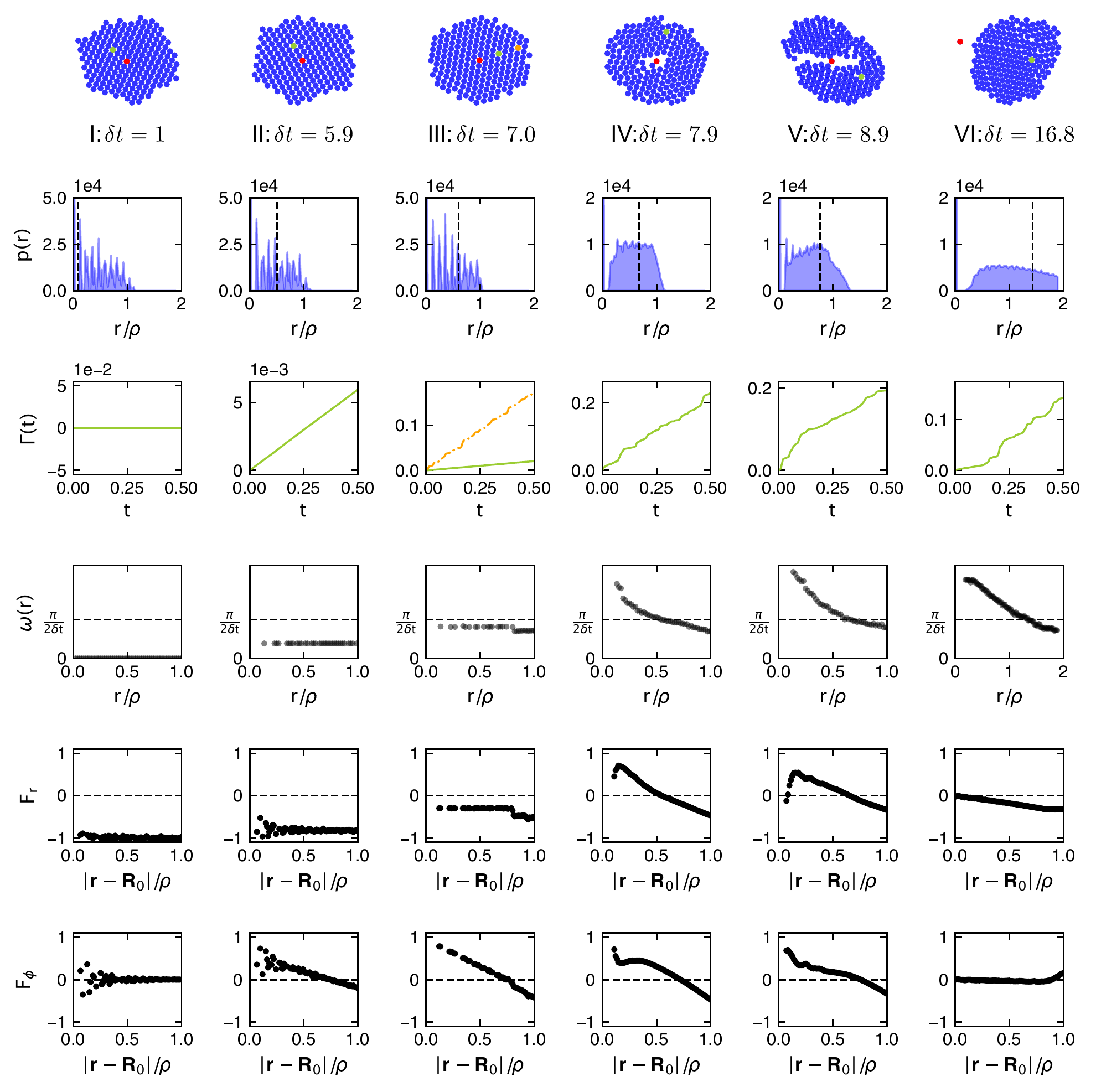}
    \caption{Crystallite configurations and their shell angular velocities $\omega(r)$ and the accumulated shear strains $\Gamma(t)$ for selected bulk particles (green and orange), as caused by the radial and tangential ``forces'' or nominal swim velocities $F_r$, $F_\phi$ in the co-moving, co-spinning frame. The corresponding force fields are shown in SM Fig.~S1. The dynamical phases I to VI of Fig.~\ref{fig:phase diagram} were simulated for vanishing thermal noise $D=0$ and $N=199$ particles (corresponding to $\rho \approx 7.43$ if close-packed). Vertical dashed lines in the radial distribution functions $p(r)$ indicate the preferred single particle orbit radius $R = 2\delta t/\pi$.}\label{fig:shell_analysis}
\end{figure*}
    
\section{Order parameters}

To distinguish between the six dynamical phases in Fig.~\ref{fig:phase diagram}, we introduce the following three order parameters. 

$\bullet$ the radial distribution $p(r)$ (the probability density to find an active particle at distance $r$ from the targeted center), normalized as $2\pi\int_0^\infty dr\, r p(r) = 1$.

$\bullet$ the (absolute) angular velocity $\omega(r)$ of concentric shells of width 0.14, given by $r^2\omega \equiv |\langle\mathbf{r}_i \times \dot{\mathbf{r}}_i \rangle|_{|\mathbf{r}_i| \approx r}$ .


$\bullet$ the cumulative shear strain $\Gamma(t) = \int_0^t dt' |\dot{\Gamma}(t')|$ around a representative bulk particle at time $t$. Formally, the shear rate is defined as $\dot{\Gamma} = (\partial{v_x}/\partial y+\partial{v_y}/\partial x)/2$, where $v_{x,y}$ denote Cartesian components of the velocity field. As a proxy for our particulate system, we use
\begin{equation}
\dot{\Gamma}(t) = \frac{1}{2} \sum_j\left(\frac{\dot{x}_i(t)-\dot{y}_j(t)}{y_i(t)-y_j(t)}+\frac{\dot{y}_i(t)-{\dot y}_j(t)}{x_i(t)-x_j(t)}\right)
\end{equation}
The sum runs over nearest-neighbor shell particles $j$ that are less than $\sqrt{2}$ away from a selected bulk particle $i$. To obtain the time derivatives of the components $x_i(t)$ and $y_i(t)$ of the position vector $\mathbf{r}_i(t)$, we average Eq.~\eqref{eq:EOM} over $200$ simulation time steps. Spurious coordinate singularities are regularized by discarding terms with denominators smaller than $0.05$.

\section{Dynamical phases}

As shown in Fig.~\ref{fig:shell_analysis}, each of the dynamical phases differs from the other five in the qualitative behavior of at least one of the characteristics $p(r)$, $\omega(r)$, and $\Gamma(t)$. The figure also shows the average radial and tangential projections of the nominal velocities (or ``forces'') $\mathbf{F}_i - \dot{\mathbf{R}}_{0}$
of particles in the co-moving frame at a given distance  from the center of mass $\mathbf{R}_{0} = \sum_{i=1}^N \mathbf{r}_i/N$ of the system. The average radial projection $F_r(r) \equiv \langle (\mathbf{F}_i - \dot{\mathbf{R}}_{0}) \cdot (\mathbf{r}_i - \mathbf{R}_{0})/r \rangle_{|\mathbf{r}_i - \mathbf{R}_{0}| \approx r}$ can be interpreted as a ``shell pressure''. In the depicted average tangential component $F_\phi(r) \equiv |\langle (\mathbf{F}_i - \dot{\mathbf{R}}_{0}) \times (\mathbf{r}_i - \mathbf{R}_{0})/r \rangle|_{|\mathbf{r}_i - \mathbf{R}_{0}| \approx r} - |\omega| r$ we also subtracted the part responsible for the crystallite's overall solid body rotation to improve the visibility of what can then be interpreted as a tangential shear stress. 
While the compression of the cluster by $F_r$ mostly helps to  maintain its crystalline structure, the tangential stress $F_\phi$ imposes unequal torques on the concentric particle shells, thereby inducing the tangential shear bands and breakup events seen in the phases III-VI. It is noteworthy that, due to the normalization of the nominal velocities $|\mathbf{F}_i| = 1$, these two competing tendencies are not independent in our system. Upon increasing the delay time $\delta t$, the nominal velocities increasingly tilt away from the central direction, meaning that the pressure decreases while the shear increases, at the same time, aggravating the destabilization. Also, for the non compact crystallites forming for longer delays, the radial forces themselves may cause radial shear bands and additionally contribute to the breaking of the crystalline configuration.

In the following, we characterize the individual dynamical phases in greater detail. The best intuitive insight into their dynamic nature is gathered from the corresponding videos in the supplementary material (SM).

\subsection{$\rm (I)$ Static crystallite: $\delta t \lesssim 0.75 \rho$}   

For short delays $\delta t$, the active Brownian particles are propelled exactly toward the target by their nominal velocities, as shown in last two lines of Figs.~\ref{fig:shell_analysis} and S1. Due to the steric repulsion, they form a non-rotating densely packed hexagonal crystallite (with small Brownian fluctuations). Its radial distribution function $p(r)$ resembles that of close-packed hard discs, while the dynamical order parameters $\omega$ and $\dot{\Gamma}$ vanish. 
    
\subsection{$\rm (II)$ Spinning crystallite: $0.75\rho < \delta t \lesssim 0.94 \rho$} 

Upon increasing the delay time $\delta t$ beyond the threshold $0.75\rho$, the crystallite exhibits solid body rotation around the target particle. The order parameters thus remain the same as in phase I, with the exception that $\omega(r) = \omega$ is given by a nonzero constant that is accurately predicted by the single-particle theory. However, as the particles' propulsion speed is fixed to 1, the particles closer to the target would individually prefer to move with larger angular velocities than those further away, while a constant $\omega(r)$ is enforced by the steric interactions and the radial pressure exerted by the particles in the periphery, which still predominantly aim at the center. These features are nicely reflected in the radial and tangential projections of the nominal velocity in Fig.~\ref{fig:shell_analysis} and the nominal velocity field in Fig.~S1 of the SM. Notice that the nominal tangential velocities of particles near the target/periphery are larger/smaller than $\omega r$, which induces the tangential shear stresses that attempt to break up the crystallite.

\begin{figure}[t!]
    \centering
    \includegraphics[trim={1.5cm 0 0 0},clip,width=1\columnwidth]{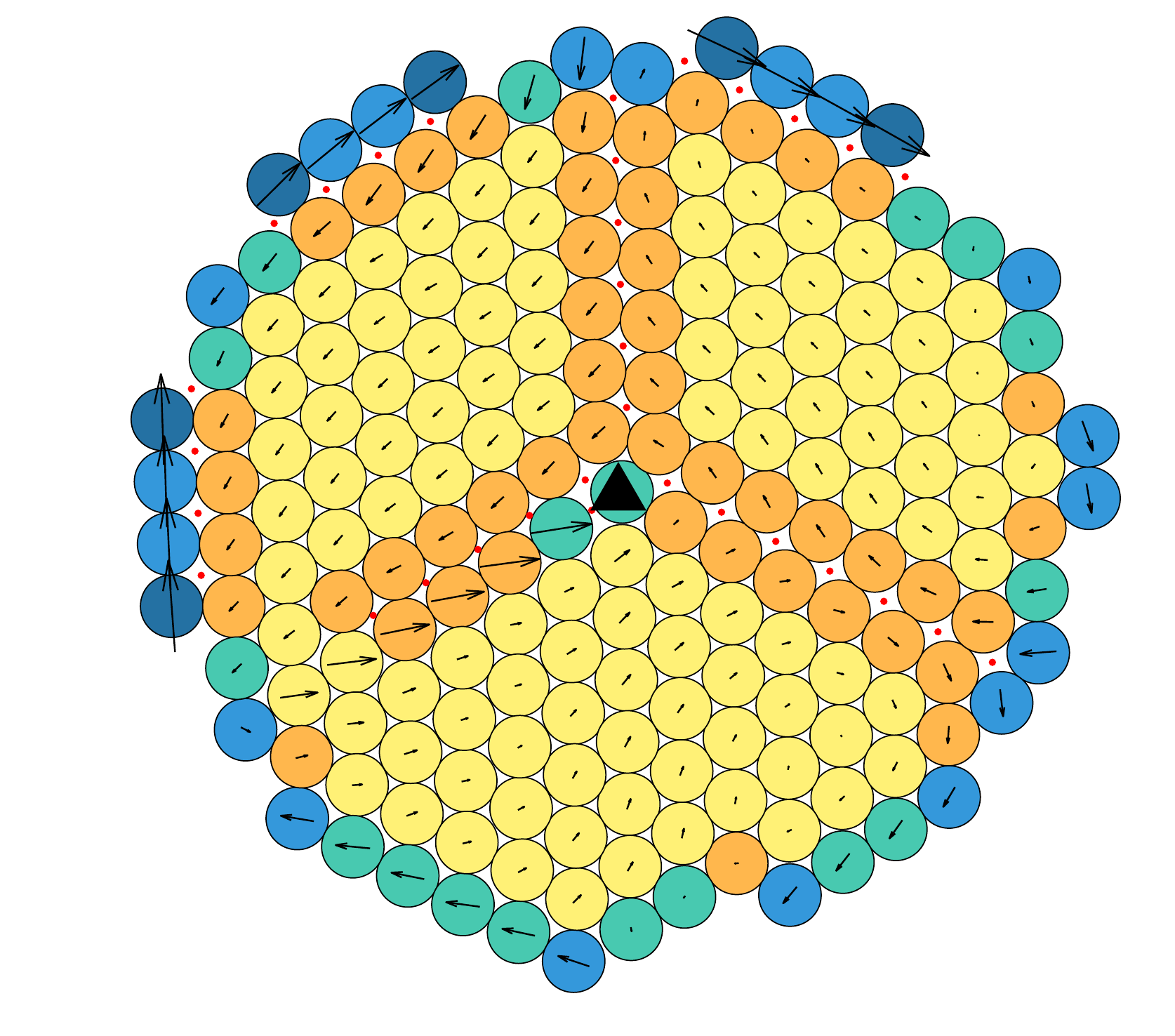}
    \caption{Snapshot of the SM video S3 ($\delta t=7.1$, $N=199$, phase III). Particle color codes for the number of nearest neighbors (from 2 to 6: steel blue, sky blue, aquamarine, orange and yellow). Red dots mark shear bands. The arrows show the actual velocities of the particles in the co-moving, co-rotating frame. The black triangle depicts the system's center of mass, which here overlaps with the central target particle. The meanings of the symbols in all videos of the SM are the same.}\label{fig:snapshot_video}
\end{figure}

\subsection{$\rm (III)$ Quaking crystallite: $0.94\rho\lesssim\delta t \lesssim 1.05\rho$}
The tangential shear stresses caused by inhomogeneous angular velocity $\omega(r)$ grow with increasing time delay. At $\delta t \approx 0.94\rho$ they overcome the compressive forces and create shear bands. As shown in Fig.~\ref{fig:shell_analysis}, the inner particles rotate (almost) at the optimal single-particle angular velocity $\pi/(2\delta t)$. The periphery lags behind, intermittently detaching and sliding around the rotating core (see the snapshots of the system in Figs.~\ref{fig:shell_analysis} and \ref{fig:snapshot_video}, and SM video 2). These stick-slip events cause a staircase-like increase of the shear strain $\Gamma(t)$ (not observed around bulk particles that are not part of a shear band), and can be interpreted as quakes of the outer shell.

The last two rows of Fig.~\ref{fig:shell_analysis} and  Fig. S1 of the SM moreover indicate that the nominal velocities of particles along radial rays from the center are no longer parallel. Closer to the center they have larger tangential components than at the periphery, creating a pressure imbalance in the system. One can interpret this as a result of the tendency of the particles to propel toward the optimal single-particle orbit, which expands with increasing $\delta t$, as indicated in the $p(r)$ panels. Upon increasing the delay somewhat beyond the value $\delta t\approx 0.94\rho$, for which the tangential shear bands appear (e.g., from $\delta t = 7$ to $\delta t = 7.1$ for $N=200$), the corresponding pressure imbalance eventually also causes the formation of system-spanning radial shear bands. Once a single radial band is formed, it destabilizes the next neighbor shell around the immobile target particle, which nucleates two more bands by the dilatancy effect, as shown in Fig.~\ref{fig:snapshot_video}. The angles between the three bands are $2 \pi/3$, corresponding to three equally sized fragments. Along the shear bands, particles slide in opposite directions (see the videos in the SM).

\subsection{$\rm (IV)$ Ring: $1.05 \rho\lesssim\delta t \lesssim 1.20\rho$}

The single particle theory predicts that the outermost layer of the crystallite would start to rotate by itself (i.e., even if the rest of the crystallite was fixed), at $\delta t = \rho$. Some of the pressure onto the crystalline core is thereupon released, which facilitates its ``breathing'' due to the dilatancy effect. The core particles can then follow more freely their tendency to approach the optimim single particle orbit, thereby creating an outward pressure (Fig.~S1). 
As a result, the crytallite detaches from the central target particle and forms a ring, in which particles inside and outside the optimal orbit converge toward it. The crystalline structure is then no longer compressed only from the outside but also from the inside.  The corresponding stresses increase the frequency of quakes and tangential and radial shear-band formation throughout the ring, as witnessed by $\Gamma(t)$ (Fig.~\ref{fig:shell_analysis}).   The associated  repeated breaking and healing effectively melts the crystalline structure  as is reflected in the monotonic decay of the angular velocity $\omega(r)$ with increasing distance $r$ from the target and the loss of structure in the radial distribution function. Both effects are somewhat moderated within the fragments forming after the permanent breakup of the ring into the yin-yang shape, described next.

\subsection{$\rm (V)$ Yin-yang/blobs: $1.2\rho\lesssim \delta t \lesssim 2.1\rho$}
 
The effective contractile force due to the inward-outward pressures described above for the ring structure destabilize the ring in a manner similar to the capillary forces in a Plateau-Rayleigh instability~\cite{Mehrabian2013}. It therefore tends to break up into $2 \pi l/2 l = \pi \approx 3$ equally sized fragments, where $2l$ is the ring width.  Due to the (essentially) athermal conditions, the exact features of the breakup depend on initial conditions, as seen in videos 8-10 of the SM. The contractile forces towards the optimal orbit also causes larger clusters to orbit more slowly than smaller ones. They contain particles further away from the optimal radius, pointing less along the orbital direction. This slows down larger fragments compared to  smaller ones, so that smaller fragments will chase the larger ones, thereby giving rise to some coarsening. 

One might therefore conclude that the many-body system would ultimately form a giant quasi-particle, centered on the optimal orbit. However, as long as the radius of the closely packed crystallite $\rho$ is larger than the optimal orbit radius $R=2\delta t/\pi$, such quasi-particle would constantly be damaged by the fixed target particle and therefore actually cannot form.  As a result, coarsening is interrupted and the system instead forms a highly dynamical yin-yang structure where the yin part continuously ``steals" particles from the yang part, and \emph{vice-versa}.
For larger delays, the yin (or the yang) component outgrows its partner until it hits the target particle. The steady state ultimately consists of a single cluster in contact with the target particle, surrounded by several sub-clusters traveling close to the optimal single-particle orbit. 
Also note that, due to their fixed speed, the particles in the fragments move with larger angular velocities  the closer they are to the center (Fig.~\ref{fig:shell_analysis}). Together with their tendency to propel towards the optimal orbit, this causes a retrograde spinning of the fragments around their own centers of mass. With respect to the order parameters depicted in Fig.~\ref{fig:shell_analysis}, the yin-yang phase exhibits the same phenomenology as the ring phase. 

To quantify the phase boundaries, we again resort to the bifurcation diagram in Fig.~\ref{fig:principle sketch}. It shows that the average angular displacement during one delay time, $\omega \delta t$, monotonically  increases with $\delta t$ up to $\delta t = 0.75 \rho \pi/2 \approx 1.18 \rho$, when it saturates at the value $\omega \delta t = \pi/2$. This is when a single active particle would detach from the fixed target particle, as its optimal orbit of radius $R$ takes off. 
This suggests that the tendency to break the ring and form a single eccentric crystallite, centered on the optimal orbit, would start at $\delta t > 1.18 \rho$, which is indeed close to the observed value $1.2\rho$, and would eventually succeed once the optimal orbit radius $R$ exceeds $\rho$. At this point a spherical crystallite would no longer interfere with fixed target particle at the center. Why this estimate fails to  provide the correct condition for the transition to the last dynamical phase is explained in the next paragraph.

\subsection{$\rm (VI)$ Satellite: $\delta t \gtrsim 2.1\rho$}

As pointed out in the preceding paragraph, one would expect to find a single compact satellite orbiting the target particle (roughly) on the  optimal single-particle orbit, when $R\approx \rho$, hence $\delta=\pi \rho/2$, which is actually not the case. The discrepancy is caused by the fact that the satellite is actually not circular but somewhat elongated along an axis that is slightly tilted relative to the radial direction. The reason is that the pressure exerted by the individual particles is no longer radially symmetric (see Fig.~S1).

The stick-slip motion of particles along the shear bands in this phase is somewhat reminiscent of an extreme version of the terrestrial tides caused by the motion of the Moon around Earth. The major difference is that the tidal forces correspond to an attraction rather than a repulsion relative to the satellite center. As a result, the  quake dynamics is approximately out of phase  by $\pi/2$, with respect to the Moon-Earth system (see video S5). Moreover, the attraction does not act toward the satellite center but toward the  optimal single-particle orbit. And finally, the elongation of the crystallite is not perfectly aligned with the direction to the center, giving rise to another phase shift that depends on the precise model parameters. 

Concerning the order parameters depicted in Fig.~\ref{fig:shell_analysis}, the satellite phase again exhibits almost the same phenomenology as the ring state. The only difference is the radial distribution of particles, which is now much broader than in the other five phases. This is indicative of the destructive effect of the tidal quakes, which  dynamically melt the crystallite into an effectively liquid droplet.

\section{Discussion and conclusions}

We have numerically studied a two-dimensional ensemble of soft active Brownian particles steered toward a target particle with a time delay. The particles form a closely packed hexagonal crystallite around the target for small delay times and experimentally relevant noise intensities. However, with increasing delay, a much richer behavior is observed. The average angular velocity around the target exhibits a bifurcation, which can be mapped to the one found recently for a single active particle~\cite{Wang2023}. An interesting ``plastic'' deformation of the hexagonal crystalline structure ensues. The tangential and radial shear stresses grow with time delay, eventually creating shear bands and breaking the crystallite. Its overall shape changes with increasing delay from a disc over a ring around the target to a yin-yang and eventually an elongated retrograde spinning satellite orbiting the target.

Our study demonstrates that simple time-delayed interactions can induce very complex dynamical behaviors in many body systems, even in the case of delayed attractions to a common fixed target. As time delays are omnipresent in interacting active matter systems in nature, this observation should be taken into account when interpreting experimental data. To this end, it would be  interesting to realize the studied many-body system experimentally. In this case, hydrodynamic interactions between the active particles would play an important role and potentially give rise to somewhat different results as obtained above, for the idealized active Brownian particle system. Our essentially athermal dynamics might thereupon become more ergodic and fluid-like~\cite{Stevens1991,Wu2009,Schall2010}. 

Our results could be extended in several other directions. First, one may consider attraction to a fixed position in space rather than to a fixed target particle. Our preliminary results with this setup reveal two major differences. Firstly, the minimum radius of rotation is determined by the noise, not by the particle diameters. Secondly, omitting the target particle increases the accessible state space. For example, the dynamics in the yin-yang/blobs phase (V) becomes much richer without the central particle, allowing for the appearance of a state with an almost deterministic periodically switching chirality. Of more practical interest might be the extension of our setup to an all-to-all attraction between the particles. Our preliminary results  show that the  phenomenology essentially remains unchanged, for short delay times. Differences appear for longer delays, where the emerging patterns are more symmetric compared to what we found above, and would deserve further study. 

\section{Acknowledgements}
\acknowledgments
We gratefully acknowledge funding through DFG-GACR cooperation by the Deutsche Forschungsgemeinschaft (DFG Project No 432421051) and by the Czech Science Foundation (GACR Project No 20-02955J). VH additionally acknowledges the support of Charles University through project PRIMUS/22/SCI/009.

\bibliography{Reference} 

\end{document}




\noindent
\begin{center}
{\LARGE\bfseries\sffamily Supplementary Material}
\end{center}

\vspace{.5cm}

\section*{\LARGE Active particles with delayed attractions form quaking crystallites}\vspace{.3cm}
\noindent
{\large\bfseries\sffamily Pin-Chuan Chen$^1$, Klaus Kroy$^1$, Frank Cichos$^2$, Xiangzun Wang$^2$ and Viktor Holubec$^{3}$}
\vspace{.25cm}
\begin{enumerate}[label={\arabic*}] 
\setlength{\itemsep}{0pt}
\item\textit{Institute for Theoretical Physics, Leipzig University, Postfach 100 902, 04009 Leipzig, Germany.}
\\
chen@itp.uni-leipzig.de, klaus.kroy@uni-leipzig.de
\item\textit{Peter Debye Institute for Soft Matter Physics, Molecular Nanophotonics Group, Universit\"at Leipzig, 04103 Leipzig, Germany.}
\\
  cichos@physik.uni-leipzig.de, wangxiangzun@gmail.com
\item\textit{Department of Macromolecular Physics, Faculty of Mathematics and Physics, Charles Univeristy, 18000 Prague, Czech Republic}
\\
  viktor.holubec@mff.cuni.cz
\end{enumerate}

\section*{Abstract}
 The supplementary information contains a figure showing the nominal velocity field for the individual dynamical phases and the description of the supplementary videos 1-10.




\vspace{0.5cm}


\setcounter{tocdepth}{2}


\section{Nominal velocity fields}

\begin{sidewaysfigure}
        \centering
        \includegraphics[width=1.1\textwidth]{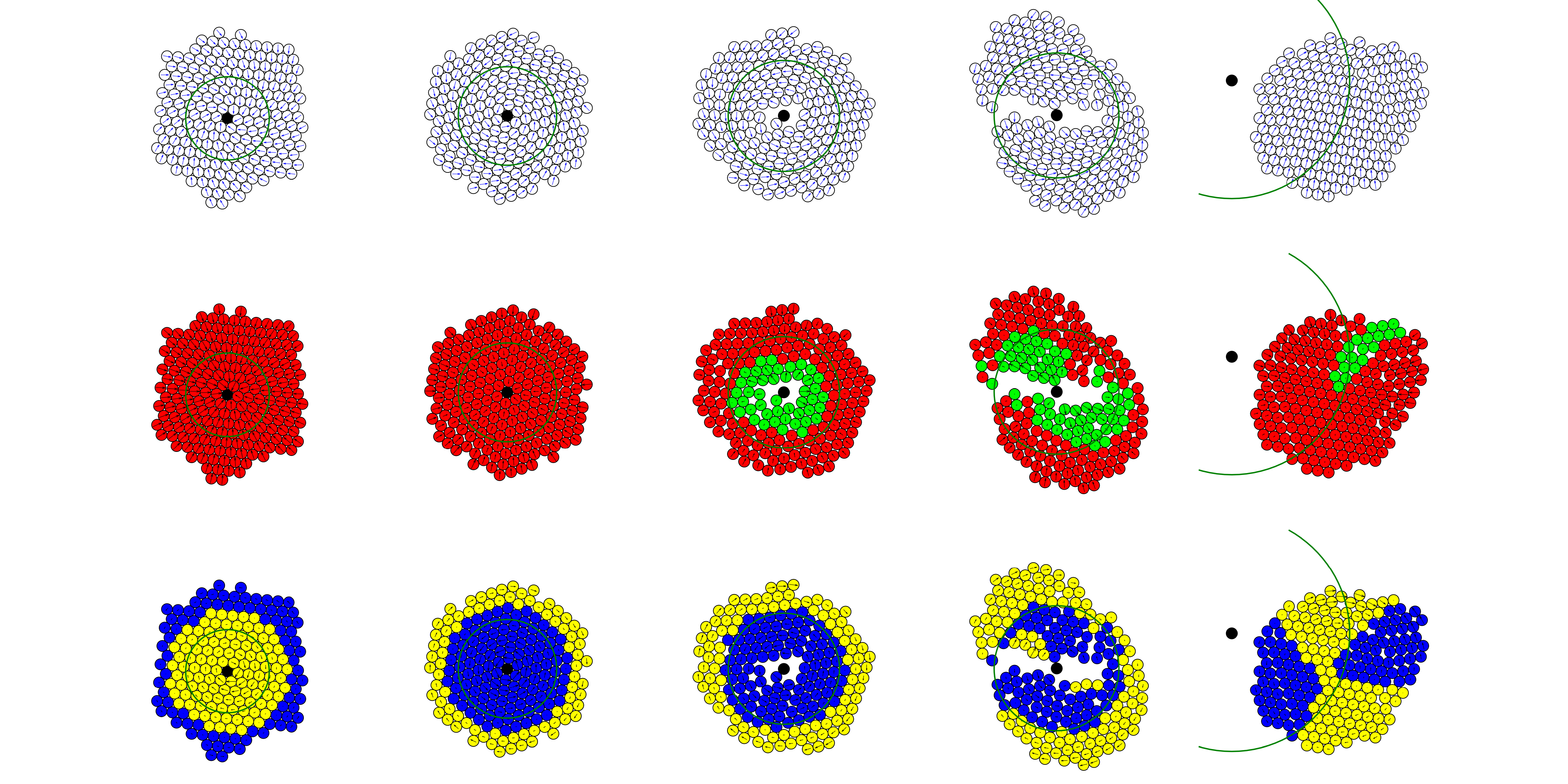}
        \caption{1st row:  nominal (intended) velocities $\mathbf{F}_i(t)$ of the individual particles in the lab frame. The whole cluster rotates in the direction of the arrows. 2nd and 3rd row: nominal velocities in the co-moving, co-spinning frame projected on the radial and tangential directions, respectively. The colors mark the directions of the arrows (red-radial from the center of mass (COM), green-radial towards the COM, yellow-clockwise rotation around COM, and blue-counter-clockwise rotation around COM). The green circles depict the optimal single particle radius $2\delta t/\pi$. The black disc indicates the fixed target particle. Delay times $\delta t$ corresponding to the individual columns are 5.9, 7, 7.9, 8.9, and 16.8, respectively. $N=200$. The averaged values of $F_r$ and $F_phi$ (2nd and 3rd row) as a function of distance to the COM are shown in the last two rows in the main text Fig. 3. In the non-rotating phase, which is not shown, the nominal velocities of all particles point to the center, as in the 1st panel of the second row.}
        \label{fig:my_label}
    \end{sidewaysfigure}

In the first row of Fig.~\ref{fig:my_label}, we show the individual particles' nominal velocities $\mathbf{F}_i$ in the lab frame. The second row of the figure depicts projections of $\mathbf{F}_i$ in the comoving, corotating frame to the radial direction from the system's center of mass:
\begin{equation}
\mathbf{F}_r^i = (\mathbf{F}_i - \dot{\mathbf{R}}_{0}) \cdot \frac{\mathbf{r}_i - \mathbf{R}_{0}}{|\mathbf{r}_i - \mathbf{R}_{0}|},
\end{equation}
where $\mathbf{R}_{0} = \sum_{i=1}^N \mathbf{r}_i/N$.
The third row of Fig.~\ref{fig:my_label} presents the tangential projections corresponding to the radial ones in the second row minus the average rotation of the system. They were calculated as 
\begin{equation}
    \mathbf{F}_\phi^i =\mathbf{F}_\parallel^i - \omega |\mathbf{r}_i - \mathbf{R}_{0}| \frac{\mathbf{F}_\parallel^i }{|\mathbf{F}_\parallel^i |},
\end{equation}
where $\mathbf{F}_\parallel^i = (\mathbf{F}_i - \dot{\mathbf{R}}_{0}) - \mathbf{F}_r^i$.

\newgeometry{left=1.6cm,bottom=0.4cm} 

    \newpage
    
    \restoregeometry

    \section{Supplementary videos}
    The particle colors in the videos code for the number of their nearest neighbors (from 0 to 6: deep blue, purple, steel blue, sky blue, aquamarine, orange, and yellow). The shear bands are marked with red dots. The arrows indicate the actual velocities of the particles in the co-moving, co-rotating frame. The black triangle depicts the center of mass of the system.

To make the shear bands better visible, videos 1-6 and 8-9 were made with zero noise ($D = 0$). Videos 7 and 10 show that nonzero noise ($D = 0.0136$) makes the dynamics of the system more erratic without changing its qualitative features. Videos 1-7 were recorded after the system reached a steady state. Videos 8-10 show the whole time
evolution of the system from the initial condition. In all the videos, we show $N=199$ particles, corresponding to $\rho \approx 7.43$. Except for the last three videos, all videos are sped up 3 times.
    
        \begin{enumerate}
        \item $\delta t=5.9$, phase II: the spinning crystallite ($D = 0$). 

        \item $\delta t= 7$, phase III: the quaking crystallite with tangential shear bands ($D = 0$).        
        

        \item $\delta t=7.1$, phase III: the quaking crystallite with tangential and radial shear bands ($D = 0$).
        
        \item $\delta t=7.9$, phase IV: the ring ($D = 0$).
        \item $\delta t=8.9$, phase V: the yin-yang/blobs ($D = 0$).
        \item $\delta t=16.8$, phase VI: the satellite ($D = 0$).
        \item $\delta t=8.9$, phase V: the yin-yang/blobs ($D = 0.0136$). 
        \item $\delta t=8.9$, phase V: the yin-yang/blobs. Typical relaxation trajectory to the yin-yang phase from a random initial condition with $D = 0$. The video is sped up 30 times.
        \item $\delta t=8.9$, phase V: the yin-yang/blobs. Another possible relaxation path to the yin-yang phase from a random initial condition with $D = 0$. The video is sped up 30 times. 
        \item $\delta t=8.9$, phase V: the yin-yang/blobs. Typical relaxation path to the yin-yang phase from a random initial condition with nonzero noise intensity $D = 0.0136$. The video is sped up 30 times.
    \end{enumerate}

